\documentclass[letterpaper, twocolumn, superscriptaddress,amsmath,amssymb,aps,prl]{revtex4-2} 

\usepackage{bbold}
\usepackage{graphicx}% Include figure files
\usepackage{dcolumn}% Align table columns on decimal point
\usepackage{bm}% bold math
\usepackage{hyperref}% add hypertext capabilities
\usepackage{braket}
\usepackage{comment}
\usepackage{xcolor}

\usepackage{orcidlink}
\newcommand{\orcidpietro}{\orcidlink{0000-0001-5279-7064}}
\newcommand{\orcidsimone}{\orcidlink{0000-0002-8882-2169}}
\newcommand{\orciddarvin}{\orcidlink{0000-0001-8805-3761}}
\newcommand{\orcidnora}{\orcidlink{0000-0002-9490-2536}}
\newcommand{\orciddaniel}{\orcidlink{0000-0001-7658-3546}}

\newcommand{\DFA}{\affiliation{Dipartimento di Fisica e Astronomia ``G. Galilei'', Università di Padova, I-35131 Padova, Italy.}}
\newcommand{\PQTC}{\affiliation{Padua Quantum Technologies Research Center, Università degli Studi di Padova}}
\newcommand{\INFNPD}{\affiliation{Istituto Nazionale di Fisica Nucleare (INFN), Sezione di Padova, I-35131 Padova, Italy.}}
\newcommand{\ULM}{\affiliation{Institute for Complex Quantum Systems, Ulm University, Albert-Einstein-Allee 11, 89069 Ulm, Germany}}

\begin{document}
\title{Entanglement transitions in a boundary-driven open quantum many-body system}

\author{Darvin Wanisch\orciddarvin}\DFA\INFNPD\PQTC
\author{Nora Reinić\orcidnora}\DFA\INFNPD\PQTC
\author{Daniel Jaschke\orciddaniel}\ULM\DFA\INFNPD\PQTC 
\author{Simone Montangero\orcidsimone}\DFA\INFNPD\PQTC
\author{Pietro Silvi\orcidpietro}\DFA\INFNPD\PQTC

\begin{abstract}
We introduce a numerical framework for integrating Markovian dynamics on tree tensor operator (TTO) ansatz states. This framework enables the simulation of both transient and steady-state regimes of systems governed by the Lindblad master equation, while preserving positivity of the density matrix and providing direct access to entanglement monotones. We demonstrate its capability to probe entanglement in open quantum many-body systems and to distinguish it from other correlations by studying a boundary-driven XXZ spin chain. Our analysis uncovers entanglement transitions driven by both the coupling to the environment and the anisotropy, revealing a striking connection between spatial entanglement scaling and spin-current.
\end{abstract}

\maketitle
Beyond its role as a vital resource for quantum technologies, entanglement is crucial for classifying complex quantum systems\,\cite{
qpt_ent_kitaev,top_order_ent_wen}, is inherently connected to their simulability by means of tensor networks\,\cite{simulability_schuch}, and is predicted to be central to foundational questions in theoretical physics\,\cite{thermalization_deutsch,thermalization_srednicki,bose_spin_ent_gravity,gravity_qi}. This significance has sparked extensive interest in studying entanglement in both equilibrium and out-of-equilibrium scenarios\,\cite{amico_rev_ent_qmb,pal_mbl_transition,tonni_neg_crit_ising,nandkishore_mbl_therm_rev,smith_dfl,skinner_mipt,rev_dynqi_lewisswan,wanisch_scrambling_lr}. Moreover, the exploration of complex quantum systems in controlled experimental setups has become increasingly feasible\,\cite{ebadi_ryd_quant_sim256,noel_meas_ind_ent_pt_ions,nishad_quant_sim_gen_spin_ex,joshi_large_scale_ent_ions,adler_2d_fragment_opt_latt}. Due to inevitable interactions with the environment, these systems are subject to dissipation, necessitating an open-system description. While dissipation is generally seen as detrimental to coherence and entanglement, given sufficient control, it can be harnessed as a resource\,\cite{harrington_eng_diss_rev}, for example, to engineer quantum many-body states\,\cite{diehl_diss_stateprep,krauter_diss_ent_gen} or perform quantum computation\,\cite{verstraete_diss_qc}. 

It is thus crucial to understand entanglement in open quantum many-body systems. However, our knowledge is largely confined to the idealized setting of closed systems, while theoretical results for open systems are scarce, and typically limited to analytically solvable models\,\cite{turkeshi_mon_bound_driv_ferm,fraenkel_free_ferm_ss_scatt,fraenkel_lr_ent_finite_temp}, or thermal quantum states\,\cite{nora_tto_ryd_therm,cocchiarella_tto_gibbs}, where entanglement follows at best an area-law\,\cite{wolf_mi_arealaw,sherman_neg_arealaw}. That said, open quantum many body systems out of equilibrium explore richer scenarios, both during the transient dynamics and in the stationary state. Potential candidates are \textit{boundary-driven open systems}\,\cite{znidaric_ent_ness,gullans_ent_curr_driv,gullans_ent_pt}, which are known to attain non-thermal stationary states\,\cite{landi_rev_bound_driv}. Furthermore, recent studies highlighted that strongly protected non-Abelian symmetries\,\cite{li_ent_ss_sym,moharramipour_sym_ent_maxmix} and Hilbert space fragmentation\,\cite{li_os_hsf} may lead to highly entangled stationary states. 

Numerical methods based on tensor networks are among the most successful frameworks in simulating both closed and open quantum many-body systems\,\cite{verstraete_mpo,vidal_eff_dyn,garcia_mps_reps,verstrate_mps_peps,schollwoeck_dmrg,orús_tn_review, daniel_os, pietro_anthology, simone_tn_book}. In closed systems, these frameworks offer straightforward access to entanglement, enabling its study out of equilibrium. However, this is not as straightforward in open systems, as currently used tensor network ans\"atze for simulating these systems\,\cite{werner_os_lptn,wellnitz_op_ent_rise_fall,preisser_op_ent_mpdo}, including approaches based on quantum trajectories\,\cite{dalibard_diss_wf_qopt,dum_mc_atomic_master,plenio_rev_qtraj}, do not provide direct access to entanglement. It was only recently shown that the \textit{tree tensor operator} (TTO)---a specific tensor network ansatz for open quantum many-body systems---grants direct access to entanglement\,\cite{arceci_tto_ent_mono}, but so far it has only been employed to thermal equilibrium\,\cite{nora_tto_ryd_therm,cocchiarella_tto_gibbs}.

In this Letter, we bridge the existing gap by developing a numerical framework that enables to probe entanglement in out-of-equilibrium open quantum many-body systems. We leverage the TTO ansatz, which allows to efficiently compute \textit{entanglement monotones}, i.e., measures that satisfy a set of properties, ensuring they exclusively probe entanglement and not other correlations\,\cite{ent_horo}. In particular, the TTO provides access to the \textit{entanglement of formation}\,\cite{eof_benett} and the \textit{logarithmic negativity}\,\cite{log_neg_plenio}. Our framework enables the study of out-of-equilibrium open quantum many-body systems in the presence of local dissipation, namely, systems governed by the \textit{Lindblad master equation} (LME) with local Lindblad operators. We apply our framework to the boundary-driven XXZ spin-chain, discovering a link between entanglement and spin-current, which allows us to connect the known transport regimes of the model to distinct entanglement scaling. 

A deeper understanding of the intricate connection between entanglement and current could enable the engineering of boundary-driven systems as a resource for entanglement generation. More broadly, our framework unlocks entirely new possibilities for exploring the uncharted territory of entanglement in open quantum many-body systems. It holds the potential to contribute to a fundamental understanding of the interplay between entanglement and dissipation, an essential pursuit in the era of noisy quantum technologies.

\begin{figure*}
    \centering
    \includegraphics[width=.98\linewidth]{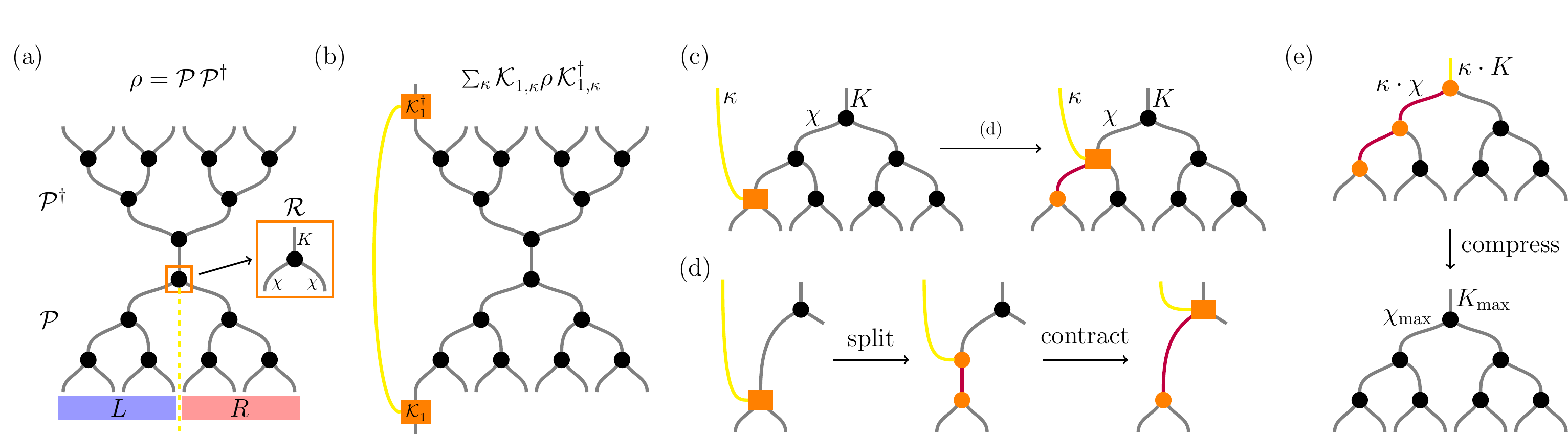}
    \caption{Tree tensor operator (TTO): a positive and loopless tensor network ansatz for density operators $\rho$. (a) The TTO consists of two branches, $\mathcal{P}$ and $\mathcal{P}^\dagger$, linked through the Kraus dimension $K$, such that $\rho = \mathcal{P} \mathcal{P}^\dagger = \sum_{k=1}^{K} p_k \ket{\Psi_k}\bra{\Psi_k}$. The root tensor $\mathcal{R}$, whose size is determined by $K$ and the bond dimension $\chi$, encodes all information on entanglement between the system's halves $L$ and $R$. Computations of entanglement monotones, such as entanglement of formation\,\cite{eof_benett} and logarithmic negativity\,\cite{log_neg_plenio}, thus only involve $\mathcal{R}$. (b) The Lindblad master equation\,\eqref{eq:lindblad_master_loc} is approximated by successively applying Eq.\,\eqref{eq:time_evo_trott}, whose unitary part is implemented via the time-dependent variational principle. The dissipative part is carried out via local Kraus channels, which are realized by connecting the Kraus tensor $\mathcal{K}_j$ to site $j$ and contracting it with its adjoint in the upper branch, as shown for $j=1$. (c) Contracting $\mathcal{K}_j$ with its connecting tensor in the TTO results in a four-leg tensor, whose additional leg is moved toward $\mathcal{R}$ via consecutive singular value decompositions (SVD), see (d). (e) After applying the Kraus channel, the bond dimension and the Kraus dimension are compressed via SVDs on the involved tensors.}
    \label{fig:num_alg_tto}
\end{figure*}
\textit{The framework.}—
An open quantum systems is described by a density operator $\rho$, representing a statistical ensemble of pure quantum states $\ket{\Psi_k}$ with corresponding probabilities $p_k$, such that $\rho=\sum_k p_k \ket{\Psi_k}\bra{\Psi_k}$. As an ansatz for $\rho$, the TTO consists of two branches: $\mathcal{P}=\sum_{k=1}^K \sqrt{p_k} \ket{\Psi_k}\bra{k}$ and its adjoint $\mathcal{P}^\dagger$, each represented as a binary tree of three-leg tensors, where the outgoing legs in the lowest (upmost) layer correspond to the local dimensions of the physical Hilbert space. The branches are linked through the Kraus dimension $K$, which determines the number of probabilities $p_k$ and respective states $\ket{\Psi_k}$ in the density operator. Consequently, the latter is given by $\rho=\mathcal{P}\mathcal{P}^\dagger$. By construction, the TTO is loopless and positive, and can be understood to arise from the global purification $\ket{\mathcal{P}}=\sum_{k=1}^K \sqrt{p_k} \ket{\Psi_k}\ket{k}$. Moreover, it has the unique feature that information on entanglement between the system's halves is contained in the uppermost tensor of $\mathcal{P}$, i.e., the \textit{root tensor} $\mathcal{R}$, provided it is the center of the unitary gauge\,\cite{arceci_tto_ent_mono,pietro_anthology}, see Fig.\,\ref{fig:num_alg_tto}\,(a). If the required bond dimension $\chi$ and Kraus dimension $K$ scale favorably with the number of degrees of freedom, the computation of entanglement monotones becomes efficient, as demonstrated for thermal quantum many-body states\,\cite{arceci_tto_ent_mono,nora_tto_ryd_therm,cocchiarella_tto_gibbs}. Generally, the TTO ansatz provides an efficient representation of the quantum state only when it is neither too entangled nor too mixed.

We embed the TTO ansatz in a non-equilibrium setting by considering systems described by the LME\,\cite{lme}
\begin{align}
    \label{eq:lindblad_master_loc}
    \dot{\rho}=i\left[\rho,\mathcal{H}\right]+\gamma\sum_{j,\alpha}\mathcal{L}_{j,\alpha}^{}\rho\mathcal{L}_{j,\alpha}^\dagger-\frac{1}{2}\left\{\mathcal{L}_{j,\alpha}^\dagger\mathcal{L}_{j,\alpha}^{},\rho\right\}\,,
\end{align}
where $\mathcal{H}$ is the system's Hamiltonian, $j$ runs over the lattice sites, and $\alpha$ over the local Lindblad operators $\mathcal{L}_{j,\alpha}$ per site, which describe the dissipative processes due to local coupling to Markovian environments. To sketch the numerical algorithm, consider that the evolution governed by Eq.\,\eqref{eq:lindblad_master_loc} for a timestep $\delta t$ is determined  by the propagator $\mathrm{e}^{\boldsymbol{\mathcal{L}}\delta t}$, where the \textit{Lindblad super-operator} $\boldsymbol{\mathcal{L}}=\boldsymbol{\mathcal{H}}+\boldsymbol{\mathcal{D}}$ consists of unitary ($\boldsymbol{\mathcal{H}}$) and dissipative part ($\boldsymbol{\mathcal{D}}$), see \textit{Endmatter} for further details. We may approximate the propagator using a symmetric second order Suzuki-Trotter decomposition\,\cite{suzuki}, 
\begin{align}
\label{eq:time_evo_trott}
\mathrm{e}^{\boldsymbol{\mathcal{L}}\delta t}=\mathrm{e}^{\boldsymbol{\mathcal{H}}\frac{\delta t}{2}}\mathrm{e}^{\boldsymbol{\mathcal{D}}\delta t}\mathrm{e}^{\boldsymbol{\mathcal{H}}\frac{\delta t}{2}}+O\left(\delta t^3\right)\,,    
\end{align}

thereby separating unitary and dissipative part. The evolution governed by Eq.\,\eqref{eq:lindblad_master_loc} is then carried out by successively applying Eq.\,\eqref{eq:time_evo_trott}, where we employ the (second order) \textit{time-dependent variational principle} (TDVP)\,\cite{haegeman_tdvp,kohn_superfluid_to_mott} for the unitary part. The assumption of local Lindblad operators further simplifies the dissipative part, yielding $\mathrm{e}^{\boldsymbol{\mathcal{D}}\delta t}=\bigotimes_j\mathrm{e}^{\boldsymbol{\mathcal{D}}_j\delta t}$. The action of each $\mathrm{e}^{\boldsymbol{\mathcal{D}}_j\delta t}$ is then realized via the corresponding local Kraus channel, $\sum_\kappa \mathcal{K}^{}_{j,\kappa}\rho\,\mathcal{K}^\dagger_{j,\kappa}$, see \textit{Endmatter} for further details. By defining the three-leg Kraus tensor $\mathcal{K}_j$, where $\left(\mathcal{K}_j\right)_\kappa=\mathcal{K}_{j,\kappa}$, applying the Kraus channel is achieved by connecting $\mathcal{K}_j^{}$ and its adjoint $\mathcal{K}_j^\dagger$ to site $j$ in the lower and upper branches, respectively, and connecting them via their Kraus leg $\kappa$, see Fig.\,\ref{fig:num_alg_tto}\,(b). Since both branches encode the same information, computations use only one; here, we focus on the lower branch. To restore the initial TTO-structure, $\mathcal{K}_j$ is contracted with its connecting tensor in the lowest layer, resulting in a four-leg tensor, see Fig.\,\ref{fig:num_alg_tto}\,(c). The additional leg of this tensor connects to the upper branch and must be moved toward the root tensor, where the two branches intersect. This is achieved by employing consecutive singular value decompositions (SVD) and contracting the tensor with the additional leg with its partner one layer above, see Fig.\,\ref{fig:num_alg_tto}(d). Applying the Kraus channel increases the bond dimension $\chi$ and the Kraus dimension $K$, which are subsequently managed in a compression step. Accordingly, we perform SVDs on the involved tensors such that $\chi$ and $K$ do not exceed the maximal values $\chi_\mathrm{max}$ and $K_\mathrm{max}$, see Fig.\,\ref{fig:num_alg_tto}(e). In summary, we obtain a second-order approximation of the LME from Eq.\,\eqref{eq:lindblad_master_loc} with additional truncation errors due to the compression steps. This algorithm is part of the \textit{QuantumTEA} library\,\cite{qtealeaves_v1_5_8}, enabling simulation of entanglement dynamics in open quantum many-body systems beyond the limits of exact diagonalization.

\begin{figure}
    %\centering
    \includegraphics[width=0.98\linewidth]{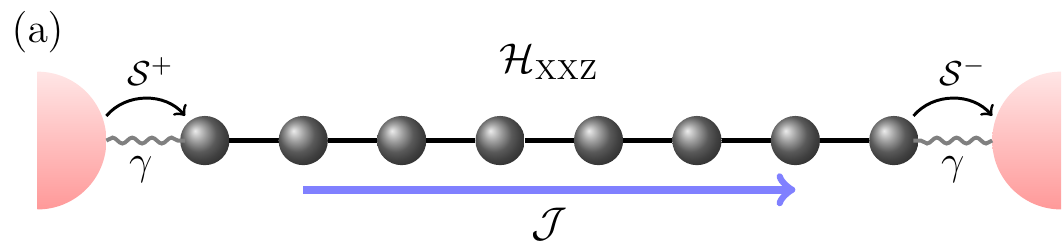}
    \includegraphics[width=0.98\linewidth]{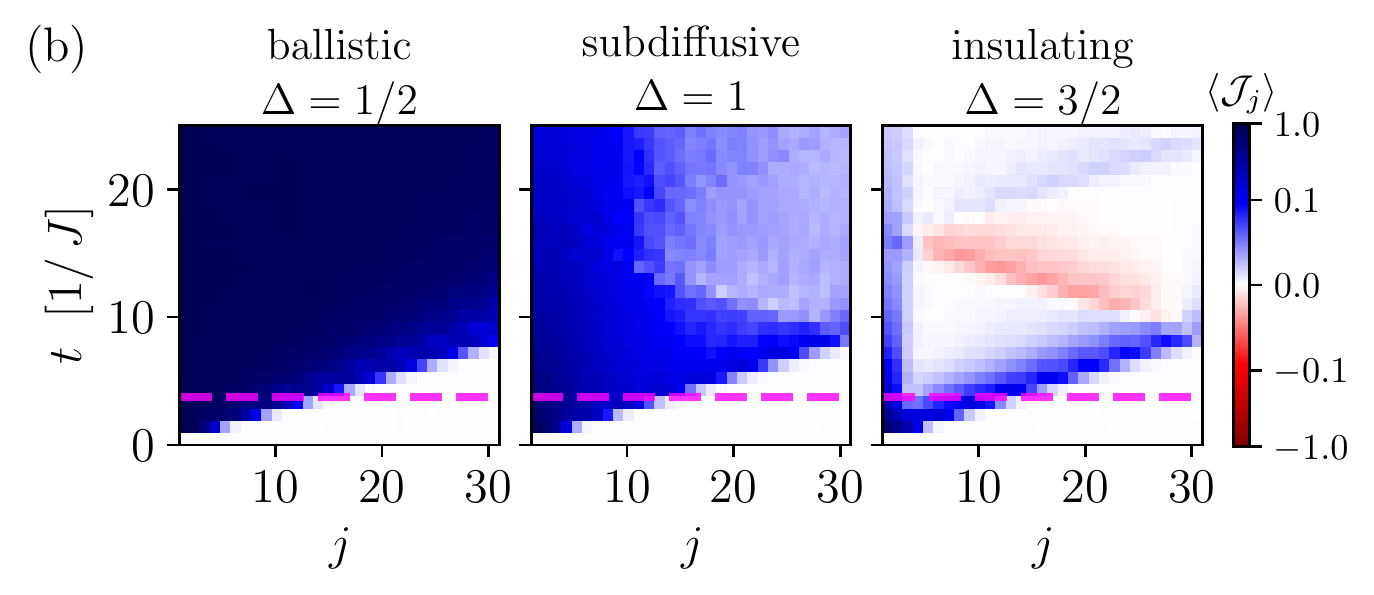}
    \caption{(a) Pictorial sketch of the boundary-driven spin-chain. The XXZ model from Eq.\,\eqref{eq:ham_xxz} is coupled at its boundaries to Markovian environments. The latter induce a current flowing through the system, resulting in a non-equilibrium stationary state. (b) Spatio-temporal profile of the spin-current $\left<\mathcal{J}_j\right>$, following a quench with initial state $\ket{Z-}$ for the different transport regimes, i.e., $\Delta=1/2,1,3/2$. System size is $\ell=32$ and $\gamma/J=1$. Horizontal dashed lines mark the arrival of the spin-current at the system's center.}
    \label{fig:pic_setup}
\end{figure}

\textit{Boundary-driven XXZ spin-chain.---}
As a paradigmatic open-system setup, we consider the boundary-driven 1D XXZ model. For $\ell$ lattice sites and open boundary conditions, its Hamiltonian is given by
\begin{align}
    \label{eq:ham_xxz}
    \mathcal{H}_\mathrm{XXZ}=\sum_{j=1}^{\ell-1}J\left(\mathcal{X}_j\mathcal{X}_{j+1}+\mathcal{Y}_j\mathcal{Y}_{j+1}+\Delta\mathcal{Z}_j\mathcal{Z}_{j+1}\right)\,,
\end{align}
where $\mathcal{X}_j,\mathcal{Y}_j,\mathcal{Z}_j$ are the Pauli operators acting on site $j$, $J>0$ is the nearest-neighbor interaction strength, and $\Delta$ is the anisotropy parameter. The system is coupled to Markovian environments at its boundaries, with Lindblad operators $\mathcal{L}_{1,1}=\sqrt{1+\mu}\,\mathcal{S}^+_1$, $\mathcal{L}_{1,2}=\sqrt{1-\mu}\,\mathcal{S}^-_1$, $\mathcal{L}_{\ell,1}=\sqrt{1+\mu}\,\mathcal{S}^-_\ell$, $\mathcal{L}_{\ell,2}=\sqrt{1-\mu}\,\mathcal{S}^+_\ell$, where $2\mathcal{S}_j^{\pm}=\mathcal{X}_j\pm i\mathcal{Y}_j$. The driving $\mu>0$ induces a spin-current flowing from left to right, resulting in a \textit{non-equilibrium stationary state}, see also Fig.\,\ref{fig:pic_setup}\,(a). This setup is commonly employed to study quantum transport\,\cite{prosen_ex_mpo_xxz1,prosen_ex_mpo_xxz2,prelovšek_easyaxis_heis_diss,nandy_perturb_heis}, as it possesses various transport regimes depending on both the anisotropy $\Delta$, and the driving $\mu$. In the following, we consider the maximally-driven case, $\mu=1$,  where analytical expressions of local observables have been found\,\cite{prosen_ex_mpo_xxz1,prosen_ex_mpo_xxz2}. Its three transport regimes, characterized by the scaling of the spin-current in the stationary state, are known as ballistic ($\Delta<1$), subdiffusive ($\Delta=1$), and insulating ($\Delta>1$)\,\cite{landi_rev_bound_driv}. In the following, we explore the non-equilibrium physics of this setup using the presented numerical framework. We refer to the \textit{Endmatter} for additional details on the numerical data.  

We initialize the system in a fully-polarized product state $\ket{Z-}=\ket{\downarrow\downarrow\downarrow\ldots}$, which is an eigenstate of the Hamiltonian from Eq.\,\eqref{eq:ham_xxz}. At $t=0$, we connect the system to the environments by quenching the coupling $\gamma$ from zero to a finite value, thereby driving the system out of equilibrium. The subsequent propagation of the spin-current $\left\langle \mathcal{J}_j\right\rangle=4\,\mathrm{Im}\left\{\left\langle\mathcal{S}^-_j\mathcal{S}^+_{j+1}\right\rangle\right\}$ is shown in Fig.\,\ref{fig:pic_setup}(b) for three values of $\Delta$, corresponding to the aforementioned transport regimes. In all cases, the early transient spin-current propagates ballistically through the system, evident from the visible linear light cone. We do not observe significant differences in the velocity of this propagation as indicated by the horizontal dashed lines, which mark the arrival of the spin-current at the center of the system. Although the transport regimes are linked to the stationary state, they also manifest in the dynamics of early to intermediate times. In the ballistic regime ($\Delta=1/2$), the spin-current rapidly saturates to a finite value as it propagates through the system. In contrast, both the subdiffusive ($\Delta=1$) and insulating ($\Delta=3/2$) regimes exhibit back propagation, which is most pronounced in the latter case, suggesting more localized dynamics.

\begin{figure}
    \includegraphics[width=0.98\linewidth]{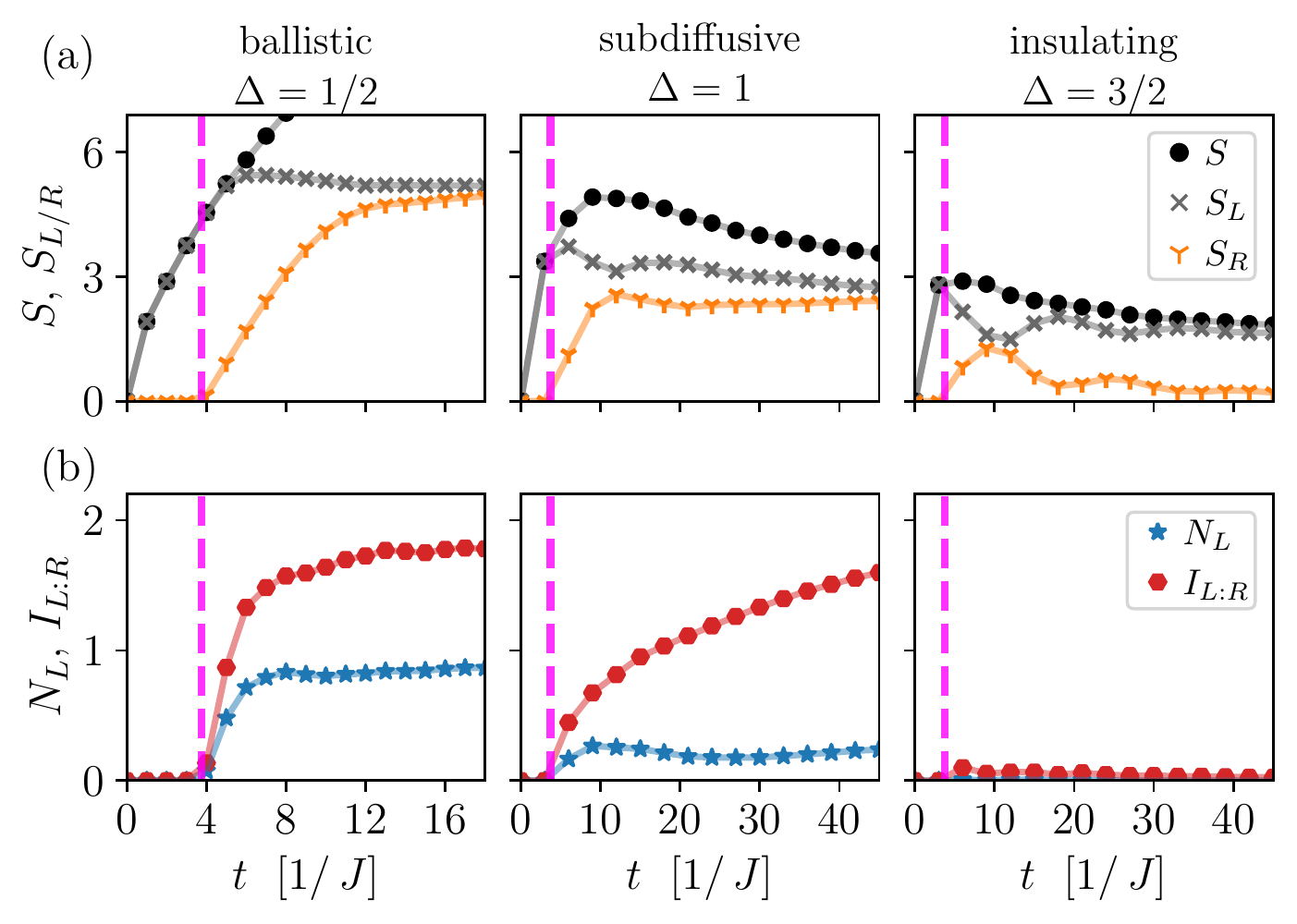}
    \caption{Dynamics following a quench with initial state $\ket{Z-}$ for the different transport regimes. System size is $\ell=32$ and $\gamma/J=1$. (a) Von Neumann entropy of the left half, $S_L$, the right half, $S_R$, and total system $S$. (b) Logarithmic negativity $N_L$ and mutual information $I_{L:R}$, quantifying the entanglement and the total correlation between left and right halves respectively. Vertical dashed lines mark the arrival of the spin-current at the system's center.}
    \label{fig:ent_growth}
\end{figure}

Our framework allows us to explore the dynamics from a different perspective by examining entropic quantities and entanglement. In Fig.\,\ref{fig:ent_growth}\,(a), we display the dynamics of the von Neumann entropy of the left half of the system, $S_L$, the right half, $S_R$, and the total system, $S$. Note that in open quantum systems, $S_L\neq S_R$ in general, unlike in closed systems. At early times, only the entropy of total system and that of the left half grow, as the Lindblad operator acting on the right boundary does not impact the initial state. As the spin-current reaches the system's center, correlations between the two halves can build up, leading to growth in the entropy of the right half. At this point, also  entanglement between the two halves can emerge, which we probe using the \textit{logarithmic negativity}\,\cite{log_neg_plenio}, defined as $N_L = \log\left(\left\|\rho^{\mathrm{T}_L}\right\|_1\right)$, where $\rho^{\mathrm{T}_L}$ is the partial transpose of $\rho$ with respect to the left half $L$, and $\left\|\cdot\right\|_1$ denotes the trace norm. The logarithmic negativity quantifies entanglement between the left and right halves. To complement this measure, we also consider the mutual information, $I_{L:R} = S_L + S_R - S$, which quantifies the total correlations between the two halves\,\cite{groisman_tot_corr}. These two quantities are displayed in Fig.\,\ref{fig:ent_growth}\,(b).

In the ballistic regime ($\Delta=1/2$), we observe a rapid growth of the logarithmic negativity $N_L$ and the mutual information $I_{L:R}$ upon the arrival of the spin-current at the system's center, followed by a quick saturation to a finite value. Both quantities exhibit similar behavior, while the logarithmic negativity confirms that a sizeable amount of correlations is dedicated to entanglement. In the subdiffusive regime ($\Delta=1$), the logarithmic negativity shows fundamentally different behavior compared to the mutual information, highlighting our framework’s ability to differentiate between entanglement from other correlations. While correlations grow monotonically, entanglement remains almost constant within the shown time interval and is lower compared to the ballistic regime. Interestingly, in the insulating regime ($\Delta=3/2$), we observe no growth of either the logarithmic negativity or the mutual information. Hence, the localized propagation of the spin-current is associated with the absence of entanglement growth. Overall, the growth of entropy, correlations, and entanglement is progressively suppressed as $\Delta$ increases, transitioning from the ballistic to the subdiffusive and finally to the insulating regime, further indicating a shift toward more localized dynamics. While this shift is also apparent from entropy analysis, access to the logarithmic negativity and the mutual information reveals a clearer distinction.

By examining the different transport regimes in more detail, we uncover distinct spatial entanglement scaling laws. As expected for a tensor-network-based approach, these scaling laws manifest in varying computational complexity across regimes. While the subdiffusive and insulating regimes allow long-time simulations with moderate bond dimension $\chi$ and Kraus dimension $K$, the ballistic regime is more demanding and limited to intermediate times and system sizes. There, the logarithmic negativity $N_L$ saturates to an extensive value, indicating a volume-law, as shown in Fig.\,\ref{fig:ent_sat}(a). Note that the system is not yet stationary at these intermediate times.

In contrast, in the subdiffusive regime we can dynamically reach the stationary state for up to $\ell=128$ sites. Figure\,\ref{fig:ent_sat}(c) shows the corresponding value of the logarithmic negativity for various system sizes, which exhibit no scaling, indicating an area-law. We also compare this behavior to the mutual information, which remains extensive, highlighting that total correlations follow a different scaling than entanglement. To further probe the entanglement structure, we compute the logarithmic negativity $N_{j,j+1}$ between pairs of neighboring sites and the rest of the system, as shown in Fig.\,\ref{fig:ent_sat}(d). This reveals local entanglement throughout most of the system, with a decay towards the boundaries.

\begin{figure}
    \includegraphics[width=0.98\linewidth]{fig_ent_scaling.pdf}
    \caption{Entanglement scaling across regimes: (a) Dynamics of logarithmic negativity $N_L$ in the ballistic regime ($\Delta = 1/2$) for various system sizes $\ell$ at $\gamma/J = 1$. (b) $N_L$ and mutual information $I_{L:R}$ at $t^* = 10J$ for $\ell = 32$, $\Delta = 1/2$, and varying $\gamma$. Solid line shows analytical spin-current in the stationary state as $\ell \rightarrow \infty$\,\cite{prosen_ex_mpo_xxz2}, exhibiting similar $\gamma$-dependence. (c) Scaling of $N_L$ and $I_{L:R}$ in the stationary state of the subdiffusive regime ($\Delta = 1$). (d) Logarithmic negativity $N_{j,j+1}$ of neighboring pairs of sites against the rest of the system in the stationary state of (c) for $\ell=128$.}
    \label{fig:ent_sat}
\end{figure}

In the insulating regime, the previously reported absence of entanglement growth persists up to the longest times practically accessible. However, the time required to reach the stationary state appears to scale exponentially with system size, making it accessible only for small system sizes. Established results for the stationary state indicate a kink-shaped magnetization profile\,\cite{prosen_ex_mpo_xxz2}, suggesting that if any entanglement is present, it would be confined to a small region in the system's center, implying at most an area-law. This is fundamentally different from the subdiffusive regime, where we find a broader distribution of entanglement, see Fig.,\ref{fig:ent_sat}(d).

Interestingly, tuning the environmental coupling $\gamma$ reinforces the observed connection between entanglement and spin-current.  In the ballistic regime ($\Delta = 1/2$), entanglement growth is strongly suppressed as $\gamma$ decreases. This is shown in Fig.\,\ref{fig:ent_sat},(b), which displays the logarithmic negativity and mutual information at fixed time $t^* = 10J$ for various $\gamma$, considering the same quench as before. The time $t^*$ is chosen such that information has already propagated past the system's center. 
Both mutual information and logarithmic negativity exhibit the same dependence on $\gamma$, aligning with the analytical value of the spin-current in the stationary state\,\cite{prosen_ex_mpo_xxz2}. Consequently, increasing $\gamma$ from small values to $\gamma \sim J$ induces a transition from vanishing to finite entanglement. Remarkably, despite considering the logarithmic negativity and mutual information at intermediate times, we observe a behavior consistent with the spin-current in the stationary state, providing additional support that the previously reported extensive scaling of logarithmic negativity in the ballistic regime also holds in the stationary state. 

\textit{Conclusions.—}
We introduced a numerical framework based on tree tensor operators, enabling large-scale simulations of out-of-equilibrium open quantum many-body systems while granting access to entanglement monotones. Leveraging this framework, we provide new insights into the boundary-driven XXZ model by uncovering a connection between entanglement and spin-current. In particular, the established transport regimes of the model correspond to distinct spatial entanglement scaling, highlighting the classificatory power of entanglement even within the realm of open quantum systems. Our results demonstrate that dissipation is necessary in this setup for achieving a sizable amount of entanglement. 

Considering future works, our framework may be employed to investigate the phase diagram of the model beyond the maximally driven case, $\mu=1$. An extended analysis could shed light on the crossover at the Heisenberg point ($\Delta=1$) from subdiffusive to superdiffusive transport as $\mu$ decreases, addressing a currently open question\,\cite{landi_rev_bound_driv}. Analyzing the robustness of highly entangled states emerging from strong non-Abelian symmetries, as proposed in Ref.\,\cite{li_ent_ss_sym}, could further clarify the feasibility of their realization with quantum technology. The framework may help assess whether current quantum technology can reach quantum advantage in exploring highly entangled states via quench experiments. By enabling controlled simulations of entanglement dynamics in open systems, it offers a platform to test hardware limits in experimentally relevant yet computationally demanding scenarios.

\textit{Acknowlegdements.---}
We thank Zala Lenarčič for fruitful discussions. The research leading to
these results has received funding from the following organizations: European Union via UNIPhD programme (Horizon 2020 under Marie Skłodowska-Curie grant agreement No.101034319 and NextGenerationEU), Italian Research Center on HPC, Big Data and Quantum Computing ICSC (NextGenerationEU Project No. CN00000013), project EuRyQa (Horizon2020), project T-NiSQ (QuantERA2021), project PASQuanS2 (Quantum Technology Flagship); Italian Ministry of University and Research (MUR) via: project TANQU (PRIN2022), Quantum Frontiers (the Departments of Excellence 2023-2027); German Federal Ministry of Education and Research (BMBF): project QRydDemo (the funding program quantum technologies - from basic research to market); the World Class Research Infrastructure - Quantum Computing and Simulation Center (QCSC) of Padova University; The italian Istituto Nazionale di Fisica Nucleare (INFN) via project Iniziativa Specifica IS-Quantum.\\ 

\textit{Endmatter.---}
Here, we provide additional details on our approach to the LME from Eq.\,\eqref{eq:lindblad_master_loc} within our numerical framework. To do so, it is useful to consider the vectorized form of the LME\,\cite{turkington}, which is given by 
\begin{align}
\label{eq:lme_vec}
\left.\left|\dot{\rho}\right>\right>=\boldsymbol{\mathcal{L}}\left.\left|\rho\right>\right>\,,\tag{E1}    
\end{align}
where $\boldsymbol{\mathcal{L}}$ is the Lindblad super-operator, consisting of unitary and dissipative part, i.e., $\boldsymbol{\mathcal{L}}=\boldsymbol{\mathcal{H}}+\boldsymbol{\mathcal{D}}$. The unitary part is given by 
\begin{align}
\boldsymbol{\mathcal{H}}=i\left(\mathbb{1}\otimes\overline{\mathcal{H}}-\mathcal{H}\otimes\mathbb{1}\right)\,,\tag{E2}   
\end{align}
and the dissipative part by $\boldsymbol{\mathcal{D}}=\gamma\sum_{j,\alpha}\boldsymbol{\mathcal{D}}_{j,\alpha}$, where
\begin{align}
    \boldsymbol{\mathcal{D}}_{j,\alpha}=
    \mathcal{L}_{j,\alpha}\otimes\overline{\mathcal{L}}_{j,\alpha}
    -\frac{1}{2}\left(\mathcal{L}_{j,\alpha}^\dagger\mathcal{L}_{j,\alpha}^{}\otimes\mathbb{1}+\mathbb{1}\otimes\mathcal{L}_{j,\alpha}^{\mathrm{T}}\overline{\mathcal{L}}_{j,\alpha}^{}\right).\tag{E3}
\end{align}
From Eq.\,\eqref{eq:lme_vec}, it follows straightforwardly that the evolution for a timestep $\delta t$ is determined by the propagator $\mathrm{e}^{\boldsymbol{\mathcal{L}}\delta t}$, as stated in the main text. Considering the decomposition of the Lindblad super-operator into unitary and dissipative part, a symmetric second-order Suzuki-Trotter decomposition\,\cite{suzuki} then yields Eq.\,\eqref{eq:time_evo_trott}. Notably, the unitary parts of two consecutive timesteps can be combined, which further reduces computational cost.

To carry out the dissipative part of Eq.\,\eqref{eq:time_evo_trott} we first exploit the assumption of local Lindblad operators, which leads us to 
\begin{align}
\label{eq:diss_part_prod}
\mathrm{e}^{\boldsymbol{\mathcal{D}}\delta t}=\bigotimes_j\mathrm{e}^{\boldsymbol{\mathcal{D}}_j\delta t}\,,\tag{E4}   
\end{align}
as discussed. Since each product in Eq.\,\eqref{eq:diss_part_prod} is a completely positive trace-preserving map\,\cite{nielsen}, there exists a set of Kraus operators $\left\{\mathcal{K}_{j,\kappa}\right\}$, such that $\mathrm{e}^{\boldsymbol{\mathcal{D}}_j\delta t}=\sum_\kappa \mathcal{K}_{j,\kappa}\otimes\overline{\mathcal{K}}_{j,\kappa}$. The matrix representation of $\mathrm{e}^{\boldsymbol{\mathcal{D}}_j\delta t}$ has dimension $d^2\times d^2$, where $d$ is the local dimension of site $j$. Obtaining the Kraus operators $\left\{\mathcal{K}_{j,\kappa}\right\}$ requires determining the eigensystem of $\mathrm{e}^{\boldsymbol{\mathcal{D}}_j\delta t}$, a procedure whose computational cost is negligible for small local dimensions $d$. Once the Kraus operators $\left\{\mathcal{K}_{j,\kappa}\right\}$ are obtained, the corresponding Kraus channel is applied as described in the main text. 

We report the maximal Kraus dimension $K_\mathrm{max}$ and maximal bond dimension $\chi_\mathrm{max}$ used to obtain the presented numerical data in Table\,\ref{tab:num}. For all results shown, we chose a timestep of $\delta t=0.025$, which is sufficient for convergence. For results regarding the system size $\ell=8$, we chose the maximal possible $K_\mathrm{max}$ and $\chi_\mathrm{max}$, meaning we consider the full Hilbert space and there is no compression error.
\begin{table}[]
    \centering
    \begin{tabular}{c|c|c|c|c|c}
        &$\Delta$&$\gamma/J$&$\ell$&$K_\mathrm{max}$&$\chi_\mathrm{max}$\\
        \hline
        Fig.\,\ref{fig:pic_setup}\,(b)&1/2&1&32&1024&128\\
        &1&1&32&512&64\\
        &3/2&1&32&512&64\\
        \hline  
          Fig.\,\ref{fig:ent_growth}\,(a,b)&1/2&1&32&1536&192\\
          &1&1&32&512&64\\
            &3/2&1&32&512&64\\
          \hline
          Fig.\,\ref{fig:ent_sat}\,(a)&1/2&1&8&256&16\\
          &1/2&1&16&1536&128\\
          &1/2&1&32&1536&192\\
          \hline
          Fig.\,\ref{fig:ent_sat}\,(b)&1/2&all shown&32&1024&160\\
          \hline
          Fig.\,\ref{fig:ent_sat}\,(c)&1&1&8&256&16\\
          &1&1&16&512&48\\
          &1&1&32&512&64\\
          &1&1&64&512&64\\
          &1&1&128&768&64\\
          \hline
          Fig.\,\ref{fig:ent_sat}\,(d)
          &1&1&64&512&64\\
    \end{tabular}
    \caption{Maximal Kraus dimension $K_\mathrm{max}$ and maximal bond dimension $\chi_\mathrm{max}$ used for the presented numerical data.}
    \label{tab:num}
\end{table}
\bibliography{ref}
\end{document}